\documentclass[12pt]{book}

\usepackage[dvips]{graphicx,color}
\usepackage{makeidx,tsukuba}
\usepackage{amssymb}

\makeauthorindex
\makeindex

\begin{document}

\BookTitle{\itshape The 28th International Cosmic Ray Conference}
\CopyRight{\copyright 2003 by Universal Academy Press, Inc.}
\pagenumbering{arabic}

\chapter{
Muon Energy Reconstruction in ANTARES and its Application to the
Diffuse Neutrino Flux}

\author{%
%
%
Alain Romeyer$^1$, Ronald Bruijn$^2$ and Juan-de-Dios Zornoza$^3$ \\
On behalf of the ANTARES collaboration\\
{\it (1) DSM/DAPNIA/SPP, CEA Saclay, 91191 Gif sur Yvette, France\\
(2) NIKHEF, Kruislaan 409, 1009 SJ Amsterdam, The Netherlands\\
(3) IFIC (CSIC-Univ. de Valencia), apdo. 22085, E-46071 Valencia, Spain}\\
}

\section*{Abstract}

The European collaboration ANTARES aims to operate a large neutrino 
telescope 
in the Mediterranean Sea, 2400 m deep, 40 km from Toulon (France). Muon 
neutrinos are detected through the muon produced in charged current 
interactions in the medium surrounding the detector. The 
Cherenkov light emitted by the muon is registered by a 3D photomultiplier 
array. Muon energy can be inferred using 3 different methods based on the 
knowledge of the features of muon energy losses.
They result in an energy resolution of a factor $\sim 2$ above 1 TeV. 
The ANTARES sensitivity to diffuse neutrino flux models is obtained from 
an energy cut, rejecting most of the atmospheric neutrino 
background which has a softer spectrum. 
Fake upgoing events from downgoing atmospheric muons are rejected using 
dedicated variables. After 1 year of data taking, the ANTARES sensitivity is  
$E^{2}~d\Phi_{\nu}/dE_{\nu} \lesssim 8 \cdot 10^{-8}~GeV~cm^{-2}~s^{-1}~sr^{-1}$ 
for a 10 string detector and an $E^{-2}$ diffuse flux spectrum.

\section{Introduction}

The ANTARES detector will
consist of 12 identical strings, each string supporting 30
storeys each of which are equipped with 3 downward looking optical modules [1].
Each optical module is a pressure resistant glass sphere containing a 10 inch
photomultiplier (PMT). The arrival times and amplitudes of the muon Cherenkov 
light hits on PMT allows the reconstruction of the track 
with a very good precision (angular resolution of 0.2$^o$ at high energy) and 
a reasonable estimation of the energy. 
In section 2, the three energy estimators developed in ANTARES are
described. In section 3, the ANTARES sensitivity to  cosmic diffuse 
neutrino fluxes are obtained after rejection of the atmospheric 
neutrino and muon background.

\section{Muon energy estimators}
\label{sec2}

The 3 energy estimators are based on the increase 
of emitted light due to muon catastrophic energy losses above 1 TeV.
The first method is purely empirical. The second is based on muon energy loss.
The third uses a neural network approach. Results are shown for a 10 string detector.

\subsection{Minimum ionising muon comparison method}
\label{subsec21}

This empirical method compares the 
recorded 
amplitude on each PMT, $A_{i}$, to the theoretical amplitude 
that a minimum ionising muon (MIM) would have given, $A_{i}^{MIM}$ [3]. 
A hit classification is determined depending on the value of the 
$\frac{A_{i}}{A_{i}^{MIM}}$ ratio for each hit {\it i}: if 
$0.1 < \frac{A_{i}}{A_{i}^{MIM}} < 100$ ($10 < \frac{A_{i}}{A_{i}^{MIM}} < 1000$) 
the hit is classsified as a ``low (high) energy''.
A variable called 
$x_{s}$, where s stands for ``low'' or ``high'', is defined for both classes: 
\begin{equation}
\label{eqx}
x_{s} = N \left(\frac{\sum_{i} A_{i}}{\sum_{i} A_{i}^{MIM}}-1\right)
\end{equation}
where i runs over all the ``low'' (``high'') energy hits 
and N is the total number of hits in the event. 
The relation between $x_{s}$ and the true Monte Carlo muon energy, 
$E_{\mu}$, is parametrised by a parabola. If 
$log_{10}(x_{low}) < 3.5$ ($log_{10}(x_{low}) > 3.5$) then the 
$x_{low} $($x_{high}$) value is used. This method gives an energy resolution
of a factor of 3 on the energy (see Fig. 1).

\subsection{Energy from an estimation of the muon dE/dx}
\label{subsec22}

The method [4] is based on the muon energy estimation from the dE/dx evolution with 
respect to its energy. 
A dE/dx estimator, 
$\rho$, is constructed using the total hit amplitude 
in the event, $\Delta A = \sum_{i} A_{i}$, the detector response R and 
the muon path length, $L_{\mu}$, in the detector sensitive volume:
\begin{equation}
\label{rho}
\rho  = \frac{\Delta A}{R ~L_{\mu}}.
\end{equation}
The sensitive volume is defined as the instrumented volume to 
which 2.5 $\lambda_{eff.~att.}$ are added, $\lambda_{eff.~att.}$ being
the effective attenuation length in the water as measured at the ANTARES site.
  
The detector response R is defined as :
\begin{equation}
\label{detector response}
R = \frac{1}{N_{pm}} \sum_{j = 1}^{N_{pm}}\frac{\alpha(\theta)}{r_{j}}~e^{\frac{-r_{j}}{\lambda_{eff.~att.}}}
\end{equation}
where $N_{pm}$ is the number of PMTs in the detector; $\alpha(\theta)$ 
is the PMT angular efficiency; 
$\frac{1}{r}~e^{\frac{-r}{\lambda_{eff.~att.}}}$ is 
the number of photons reaching the PMT after a distance r in water.
The ratio $\frac{\Delta A}{R}$ is a direct estimation of the total amount 
of energy which has been lost by the muon in this volume.
The relation between $\rho$ and the true muon energy $E_{\mu}$ is fitted 
by a third order polynomial. This method is valid only when muons have a 
path length $\gtrsim 200~m$ in the sensitive volume and is
relatively insensitive to the detector geometry. The achieved energy 
resolution is a factor 2 on the muon energy above 10 TeV (see Fig. 1). 

\subsection{Neural network approach}
\label{subsec23}

This method essentially uses as input the amplitude of hit and time difference
with respect to a pure minimum ionising muon. Catastrophic muon energy 
losses lead to greater amplitudes and broader time distributions. 
The only output variable is the muon estimated energy. Actually, the NN was trained
only with events $\ge$ 1 TeV. Works is in progress for lower energies. 
\begin{figure}
  \begin{center}
    \includegraphics[height=12pc]{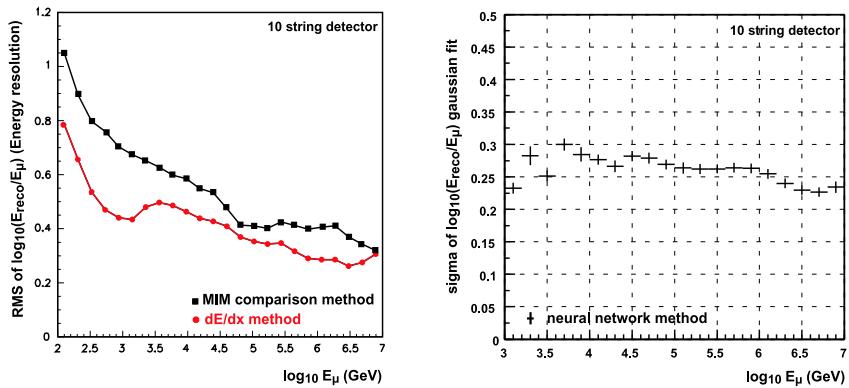}
 \end{center}
  \vspace{-0.5pc}
  \caption{Comparison of the energy estimator performances as a function of 
the true muon energy $E_{\mu}$. Left plot: evolution of the $log_{10}(\frac{E_{reco}}{E_{\mu}})$ 
RMS (square: MIM comparison method, Sec. 2.1; circle: dE/dx method, 
Sec. 2.2) [4]; right plot: evolution of the sigma of a 
gaussian fit on the $log_{10}(\frac{E_{reco}}{E_{\mu}})$ distribution 
(neural network approach, Sec. 2.3).}
\label{fig:ereco}
\end{figure}
First results show that
between 100 GeV and 1 TeV the accuracy is around a factor of 3.
Above 10 TeV, the performances are comparable to those presented in 2.2 
(around a factor 2 on $E_{\mu}$, see Fig. 1). 

\section{ANTARES sensitivity to a diffuse high energy cosmic neutrino flux}
\label{sec3}

The ANTARES sensitivity to a diffuse cosmic neutrino flux superimposed 
on top of the atmospheric neutrino background can be evaluated using 
the 3 energy estimators. For brevity, only the results of the estimator 
described in Sec.~2.2 are presented. A typical cosmic differential 
neutrino spectrum with $E^{-2}$ energy dependence has first been 
considered [5]. The atmospheric neutrino spectrum ($\sim E^{-3.6}$) is 
softer and is rejected using an energy cut, which also rejects those 
atmospheric muon background events surviving dedicated muon rejection cuts. 
These cuts are essentially independent of the assumed spectral shape.
The ANTARES sensitivity, with a 10 string 
detector, is found to be
$7.8 \pm 0.99 \cdot 10^{-8}~GeV~cm^{-2}~s^{-1}~sr^{-1}$ after 1 year 
of data taking ($3.9 \pm 0.7 \cdot 10^{-8}$ $GeV~cm^{-2} s^{-1} sr^{-1}$ after 
3 years) with an optimal 
muon energy threshold at 50 TeV (126 TeV). The uncertainties are 
determined from various predictions of the 
atmospheric prompt neutrino flux [2] (see Fig. 2). 
Using the same ``experimental'' conditions (10 string detector, 
1 year of acquisition), the sensitivity to a model like SDSS91 (M95 loud A) 
is $\sim$20 times better ($\sim$67 times worse) 
than the model flux with an optimal energy threshold at 125 TeV (210 TeV)
(see Fig. 2) [4].  
\begin{figure}
  \begin{center}
    \includegraphics[height=16pc]{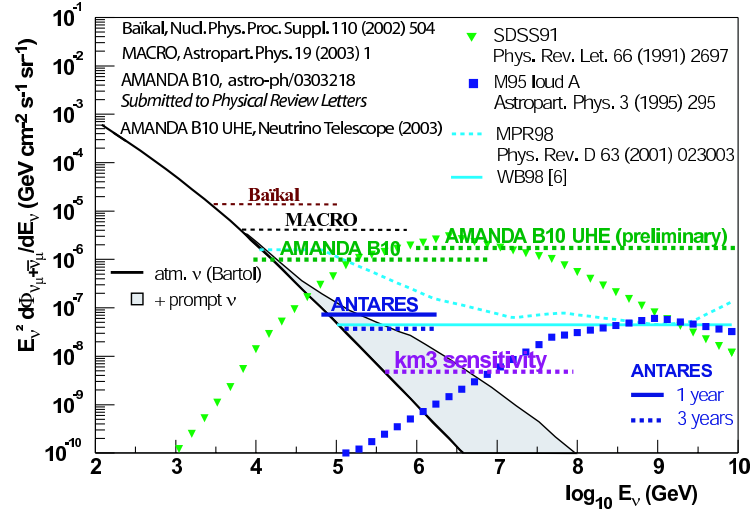}
 \end{center}
  \vspace{-0.5pc}
  \caption{ANTARES sensitivity compared to diffuse fluxes and experimental 
limits [5]. The limit is shown upon an $E^{-2}$ hypothesis.}
\label{fig:limit}
\end{figure}

\section{Conclusion}

Three different energy estimators have been developed in the ANTARES 
collaboration. They reach an energy resolution of a factor between 2 and 3 
on $E_{\mu}$ and allows a limit of 
$\sim 8 \cdot 10^{-8}~GeV~cm^{-2}~s^{-1}~sr^{-1}$ to be set 
for 1 year's data with an $E^{-2}$ spectrum and a 10 string detector. 

\vspace{\baselineskip}
\re
1.\ E. Aslanides et al.\ 1999, ANTARES proposal astro-ph/9907432
\re
2.\ C.G.S. Costa\ 2001, Astropart. Phys. 16(193)
\re
3.\ A. Oppelt\ 2001, PhD thesis, Universit\'e de la M\'editerran\'ee
\re
4.\ A. Romeyer\ 2003, PhD thesis, Universit\'e Paris VII
\re
5.\ E. Waxman and J. Bahcall\ 1999, Phys. Rev. D 59(0203002)

\endofpaper
\end{document}